\documentstyle[12pt,bezier]{article}
\begin{document}

\def \inbar{\vrule height1.5ex width.4pt depth0pt}
\def \xC{\relax\hbox{\kern.25em$\inbar\kern-.3em{\rm C}$}}
\def \xR{\relax{\rm I\kern-.18em R}}
\newcommand{\xZ}{Z \hspace{-.08in}Z}
\newcommand{\xbe}{\begin{equation}}
\newcommand{\xee}{\end{equation}}
\newcommand{\xbea}{\begin{eqnarray}}
\newcommand{\xeea}{\end{eqnarray}}
\newcommand{\xnn}{\nonumber}
\newcommand{\xkt}{\rangle}
\newcommand{\xbr}{\langle}
\newcommand{\cun}{{\mbox{\tiny${\cal N}$}}}

\title{Adiabatic Product Expansion}
\author{Ali Mostafazadeh\thanks{E-mail: alimos@phys.ualberta.ca}\\ \\
Theoretical Physics Institute, University of Alberta, \\
Edmonton, Alberta,  Canada T6G 2J1.}
\date{June 1996}
\maketitle

\begin{abstract}
The time-evolution operator for an explicitly time-dependent
Hamiltonian is expressed as the product of a sequence of 
unitary operators. These are obtained by successive time-dependent
unitary transformations of the Hilbert space followed by the 
adiabatic approximation at each step. The resulting adiabatic
product expansion yields a generalization of the quantum
adiabatic approximation. Furthermore, it leads to an infinite
class of exactly solvable models.
\end{abstract}

\vspace{.3cm}
\baselineskip=18pt

Consider  the dynamics of a quantum mechanical system whose 
Hamiltonian $H=H(\tau)$ is explicitly time-dependent. The evolution
of a state vector $|\psi(\tau)\xkt$ is governed by the Schr\"odinger
equation:
	\xbe
	i\hbar|\dot\psi(\tau)\xkt=H(\tau)\:|\psi(\tau)\xkt\;,~~~~~
	|\psi(0)\xkt=|\psi_0\xkt\;,
	\label{sch-eq}
	\xee
where a dot means a time-derivative.
Alternatively, one has $|\psi(\tau)\xkt=U(\tau)|\psi_0\xkt$, where
	\xbe
	U(\tau)={\cal T}\:e^{-\frac{i}{\hbar}\int_0^\tau  H(t)\:dt}
	\label{ev-op}
	\xee
is the time-evolution operator. Here ${\cal T}$ denotes the time-ordering
operator. The purpose of this article is to express $U(\tau)$ as the product of
a sequence of unitary operators each of which is the adiabatically
approximate time-evolution operator in some adiabatically moving frame.

In order to construct these unitary operators, let us separate the adiabatic
part  $U^{(0)}(\tau)$ of the exact time-evolution operator:
	\xbe
	U(\tau)=U^{(0)}(\tau)\: V^{(1)}(\tau)\;.
	\label{u=uv}
	\xee
$U^{(0)}(\tau)$ is defined by
	\xbea
	U^{(0)}(\tau)&:=&\sum_ne^{i\alpha_n(\tau)}\, |n;\tau\rangle\xbr n;0|\;,
	\label{u0}\\
	\alpha_n(\tau)&:=&\delta_n(\tau)+\gamma_n(\tau)\,,
	\label{al}\\
	\delta_n(\tau)&:=&-\frac{1}{\hbar}\int_0^\tau E_n(t)dt\,,~~~~
	\gamma_n(\tau)\: :=\: i\int_0^\tau\langle n;t|\frac{d}{dt}|n;t\rangle\:dt\;,
	\label{de-ga}
	\xeea
where  $|n;t\rangle$ are instantaneous eigenvectors of the
Hamiltonian $H(t)$ with eigenvalue $E_n(t)$, i.e.,
	\begin{equation}
	H(t)|n;t\rangle=E_n(t)\,|n;t\rangle\;.
	\label{ei-va-eq}
	\end{equation}
Throughout this article, it is assumed that the energy spectrum is
discrete, all the eigenvalues of the Hamiltonian are  non-degenerate
and there is no level-crossing. 

The adiabatic time-evolution operator $U^{(0)}(\tau)$ as defined by
(\ref{u0}) includes the effects of Berry's geometric phase
\cite{berry1984}, through the phase angles $\gamma_n(t)$. The operator
$V^{(1)}(\tau)$ includes the non-adiabatic effects. As $U^{(0)}(\tau)$ is
unitary, Eq.~(\ref{u=uv}) may be viewed as the definition of $V^{(1)}(\tau)$.

Next consider an arbitrary time-dependent unitary transformation of
the state vectors: $|\psi(\tau)\xkt\to|\psi'(\tau)\xkt:={\cal U}(\tau)|
\psi(\tau)\xkt$. This may be viewed as transforming to a ``moving frame''
of reference. In the ``moving frame'' the state vectors also satisfy a 
Schr\"odinger equation with a new Hamiltonian ${\cal H}(\tau)$. This is
related to $H(\tau)$ according
to
	\xbe
	{\cal H}(\tau)={\cal U}(\tau)\: H(\tau)\: {\cal U}^\dagger(\tau)-i\hbar\:
	{\cal U}(\tau)\: \dot{\cal U}^\dagger(\tau)\;.
	\label{h-tr}
	\xee
Now let us set ${\cal U}(\tau)=U^{(0)\dagger}(\tau)$. Then one can easily 
check that the state vectors $|\psi'(\tau)\xkt$ in the ``adiabatic moving 
frame'' evolve according to $|\psi'(\tau)\xkt=V^{(1)}(\tau)|\psi'(0)\xkt=
V^{(1)}(\tau)|\psi_0\xkt$. Thus, in view of Eq.~(\ref{h-tr}), we can express
$V^{(1)}(\tau)$ directly in terms of the corresponding transformed 
Hamiltonian which we shall denote by $H^{(1)}(\tau)$, namely:
	\xbe
	V^{(1)}(\tau)={\cal T}\: e^{-\frac{i}{\hbar}\int_0^\tau H^{(1)}(t)\:dt}\;.
	\label{v1}
	\xee
Since $V^{(1)}(\tau)$ is also a time-evolution operator, the same procedure
may be repeated for $V^{(1)}(\tau)$. In other words, we can use $H^{(1)}(\tau)$
in place of the original Hamiltonian $H(\tau)$ to define the adiabatic part
$U^{(1)}(\tau)$ of the time-evolution operator $V^{(1)}(\tau)$ and therefore write
$V^{(1)}(\tau)=U^{(1)}(\tau)V^{(2)}(\tau)$. Clearly this can be continued 
indefinitely. The result is an {\em adiabatic product expansion} of the 
time-evolution operator:
	\xbe
	U(\tau)=\prod_{i=0}^\infty U^{(i)}(\tau)\;.
	\label{ad-pr-ex}
	\xee

This expression may be viewed as a generalization of the quantum
adiabatic approximation \cite{messiah}. This generalized quantum 
adiabatic approximation is carried out by calculating the first $N$
terms in the product  in (\ref{ad-pr-ex}) and approximating $U(\tau)$ by
	\xbe
	\prod_{i=0}^N U^{(i)}(\tau)\;.
	\label{ge-qu-ad-ap}
	\xee
Incidentally the transformed Hamiltonian $H^{(1)}(\tau)$ may be easily
obtained in the eigenbasis of the initial Hamiltonian. This is done by 
substituting $U^{(0)\dagger}(\tau)$ for ${\cal U}(\tau)$ in (\ref{h-tr}). The
result is
	\xbea
	H^{(1)}(\tau)&=&-i\hbar\sum_{m\neq n}e^{-i[\alpha_m(\tau)-
	\alpha_n(\tau)]}\: A_{mn}(\tau)\: |m;0\xkt\xbr n;0|\;,\xnn\\
	&=&
	i\hbar\sum_{m\neq n}e^{-i[\alpha_m(\tau)-
	\alpha_n(\tau)]}\:\frac{\xbr m;\tau|\dot H(\tau)
	|n;\tau\xkt}{E_m(\tau)-E_n(\tau)}\:|m;0\xkt\xbr n;0|\;,
	\label{h1}
	\xeea
where $A_{mn}(\tau):=\xbr m;\tau|\frac{d}{d\tau}|n;\tau\xkt$, and the
second equality is obtained by differentiating Eq.~(\ref{ei-va-eq})
with respect to time and using orthonormality of the energy 
eigenvectors. The appearance of the time-derivative of the Hamiltonian
in the expression for $H^{(1)}(\tau)$ and the fact that  $H^{(1)}(\tau)$ is
off-diagonal in the eigenbasis of $H(0)$ are reminiscent of the meaning
of the adiabatic approximation in which $H^{(1)}(\tau)$ is neglected. 

Eq.~(\ref{h1}) can also be used  to yield the transformed Hamiltonian
$H^{(i+1)}(\tau)$ corresponding to the $(i+1)$-th term in (\ref{ad-pr-ex}). This
is done by introducing the symbol $n^{(i)}$ which labels the eigenvectors
and eigenvalues of  $H^{(i)}(\tau)$, and replacing $H^{(1)}(\tau)$,  $n,~m$,
$H(\tau)$ in (\ref{h1}) by $H^{(i+1)}(\tau)$, $n^{(i)},~m^{(i)}$, and $H^{(i)}(\tau)$,
respectively.  Clearly, $H^{(i)}(\tau)$ and consequently 
	\xbe
	U^{(i)}(\tau):=\sum_{n^{(i)}}e^{i\alpha_{n^{(i)}}(\tau)}\, 
	|n^{(i)};\tau\rangle\xbr n^{(i)};0|\;,
	\label{ui}
	\xee
involve $i$-th time derivatives of the original Hamiltonian. 

The generalized quantum adiabatic approximation (\ref{ge-qu-ad-ap})
of order $N$ is a reliable approximation for the time-evolution operator,
if one can neglect $H^{(N+1)}(\tau)$. It is exact, if  $H^{(N+1)}(\tau)=0$.
In fact, the latter equation may be used as a defining condition for
generating a class of exactly solvable examples for which
$U(\tau)=U^{(0)}(\tau)\cdots U^{(N)}(\tau)$ is the exact time-evolution
operator. 

In the remainder of this article, I shall demonstrate the application of 
the adiabatic product expansion in the analysis of the dynamics of
a magnetic dipole in a  changing  magnetic field. 

The Hamiltonian of this system is given by
	\begin{equation}
	H[R]=H(r,\theta,\varphi)=b \vec R(r,\theta,\varphi)\cdot \vec J
	=br(\sin\theta\cos\varphi J^1+\sin\theta\sin\varphi J^2+
	\cos\theta J^3)\:,
	\label{3.1}
	\end{equation}
where $b$ is the Larmor frequency,  $(r,\theta, \varphi)$ are spherical
coordinates and $\vec J$ is the angular momentum operator with 
components $J^\mu$, $\mu=1,2,3$. The condition $r\neq 0$ guarantees
the lack of level crossing. Therefore the
appropriate parameter space $M$ of this system is $\xR^3-\{0\}$. The 
time-dependence of the Hamiltonian corresponds to the curve $C:[0,T]\to M$
traced by the tip of the magnetic field in time. This curve is parametrically
given by $(r(t),\theta(t),\varphi(t))$. Assuming
that for all $t\in [0,T]$, $C(t)\neq (r, \theta=\pi,\varphi)$, one can choose
a complete orthonormal set of single-valued  eigenvectors $|n;R\xkt$ of 
$H[R]$. These are given by \cite{bohm-qm}:
	\xbe
	|n;(r,\theta,\varphi)\rangle=|n;(r_0,\theta,\varphi)\rangle=
	W(\theta,\varphi)
	|n;(r_0,0,0)\rangle\;, ~~~~~\theta\in[0,\pi),~\varphi\in[0,2\pi)\;,
	\label{eg-ve}
	\xee
where 
	\xbe
	W(\theta,\varphi):=e^{-\frac{i\varphi}{\hbar}J^3}
	e^{-\frac{i\theta}{\hbar}J^2}e^{\frac{i\varphi}{\hbar}J^3}\;,~~~
	(r_0,\theta_0,\varphi_0):=(r(0),\theta(0),\varphi(0)).
	\label{w}
	\xee
The corresponding eigenvalues are:
	\xbe
	E_n(r,\theta,\varphi)=E_n(r,0,0)=\hbar \,nbr\:,~~~{\rm with}~~
	n=0,\:\pm\frac{1}{2}\,,\:\pm1\,,\:\pm\frac{3}{2}\,,\cdots\;.
	\label{eg-va}
	\xee
Note that in (\ref{eg-ve}),  $|n;(r,0,0)\rangle$ are the eigenvectors of 
$H(r,\theta=0,\varphi=0)=brJ^3$, i.e., 
	\[ J^3|n;(r,0,0)\rangle=\hbar n|n;(r,0,0)\rangle \;.\]

Assuming that the eigenvectors $|n;(r,\theta,\varphi)\rangle$ of the
Hamiltonian are also eigenvectors of  $|\vec J|^2$, i.e., they have
definite angular momentum:
	\xbe
	|\vec J|^2\:|n;(r,\theta,\varphi)\rangle=j(j+1)\:
	|n;(r,\theta,\varphi)\rangle\;,~~~~ n=-j,-j+1,\cdots,j\;,
	\label{j2}
	\xee
one can calculate
	\xbea
	A_{mn}(t)&=&A_r^{(mn)}\:\dot r(t)+A_\theta^{(mn)}\:\dot \theta(t)
	+A_\varphi^{(mn)}\:\dot\varphi(t)
	\label{a}\\
	A_r^{(mn)}&:=&\langle m;(r,\theta,\varphi)|\frac{\partial}{\partial r}
	|n;(r,\theta,\varphi)\rangle\:=\: 0\;,
	\label{a_r}\\
	A_\theta^{(mn)}&:=&\langle m;(r,\theta,\varphi)|
	\frac{\partial}{\partial\theta}
	|n;(r,\theta,\varphi)\rangle\:=\:\frac{1}{2}\:(e^{i\varphi}
	C_m\delta_{m\,n-1}-e^{-i\varphi}C_n\delta_{m-1\,n})\;,
	\label{a_theta}\\
	A_\varphi^{(mn)}&:=&\langle m;(r,\theta,\varphi)|
	\frac{\partial}{\partial\varphi}
	|n;(r,\theta,\varphi)\rangle\;,\nonumber\\
	&=&i\left[ m(1-\cos\theta)\delta_{mn}+
	\frac{1}{2}\: \sin\theta\: (e^{i\varphi}C_m\delta_{m\,n-1}+
	e^{-i\varphi}C_n\delta_{m-1\,n})\right]\;,
	\label{a_phi}\\
	H^{(1)}(t)&=&  \frac{1}{2}\: W(\theta_0,\varphi_0) \left[ 
	\Omega(t)\,J^+ + \Omega^*(t)\,J^-\right] W^\dagger
	(\theta_0,\varphi_0)\;,
	\label{h1-1}\\
	\delta_n(t)&=&-nb \int_0^t r(t')\:dt'\;,~~~~~~~
	\gamma_n(t)\:=\:-n\gamma(t)\;,
	\label{de-ga-1}
	\xeea
where $C_m:=\sqrt{(j-m)(j-m+1)}$, $W$ is defined in Eq.~(\ref{w}),
	\xbea
	\Omega(t)&:=&e^{-i[\alpha(t)+\varphi(t)]}\:[\sin\theta(t)\:
	\dot\varphi(t)+i\dot\theta(t)]\,,~~~~\alpha(t)\::=\:\delta(t)+\gamma(t)\,,
	\label{v}\\
	\delta(t)&:=&-b\int_0^t r(t')\:dt'\,,~~~~
	\gamma(t)\::=\:-\int_0^t[1-\cos\theta(t')]\,\dot\varphi(t')\:dt'\,,~~
	\label{ga}
	\xeea
and extensive use is made of the properties of $J^\mu$ and $J^\pm:=
J^1\pm i J^2$, particularly:
	\[J^\pm\,|m;(r,0,0)\xkt=\hbar\,C_{\pm m}\:|m\pm 1;(r,0,0)\xkt\;.\]

Substituting (\ref{w}) and (\ref{v}) in (\ref{h1-1}), and carrying out the
necessary computations, one obtains:
	\xbea
	H^{(1)}(t)&=&\omega(t)\:\left\{ 
	[\cos^2\frac{\theta_0}{2}\cos\sigma(t)-\sin^2\frac{\theta_0}{2}
	\cos(2\varphi_0+\sigma(t))]\,J^1+\right.\xnn\\
	&&\left. [-\cos^2\frac{\theta_0}{2}\sin\sigma(t)-
	\sin^2\frac{\theta_0}{2}
	\sin(2\varphi_0+\sigma(t))]\,J^2+[-\sin\theta_0\cos\sigma(t)]\,
	J^3\right\}\;,
	\label{h1-2}\\
	&=:&b\,r^{(1)}(t)\left[\sin\theta^{(1)}(t)\cos\varphi^{(1)}(t)\, J^1+
	\sin\theta^{(1)}(t)\sin\varphi^{(1)}(t)\, J^2+
	\cos\theta^{(1)}(t)\, J^3\right],~~~
	\label{h1-3}
	\xeea
where $\omega(t):=|\Omega(t)|=\sqrt{\dot\theta^2+\sin^2\theta\,
\dot\varphi^2}$ is the angular speed of the tip of the magnetic field, 
$\sigma(t)$ is the phase of $\Omega(t)$, i.e.,
	\[ \sigma(t):=-\alpha-\varphi+\xi~~~~~~{\rm mod}~2\pi\,,\]
	\[\cos\xi:=\frac{\sin\theta\:\dot\varphi}{\omega}\,,~~~~
	\sin\xi:=\frac{\dot\theta}{\omega}\,,\]
and $r^{(1)}(t), \theta^{(1)}(t),\varphi^{(1)}(t)$
are defined by equating the right hand sides of  (\ref{h1-2}) and 
(\ref{h1-3}), i.e.,
	\xbea
	r^{(1)}(t)&:=&\frac{\omega(t)\Delta(t)}{b}\:,~~~~\Delta(t)\::=\:\sqrt{
	1+\sin\varphi_0\sin^2\theta_0\sin[\varphi_0+2\sigma(t)]}\;,
	\label{r1-De}\\
	\theta^{(1)}(t)&:=&\cos^{-1}\left[-\frac{\sin\theta_0
	\cos\sigma(t)}{\Delta(t)}	\right]\;,
	\label{theta1}\\
	\varphi^{(1)}(t)&:=&\tan^{-1}\left[\frac{\sin\sigma(t)+
	\tan^2(\frac{\theta_0}{2})\sin[2\varphi_0+\sigma(t)]}{
	-\cos\sigma(t)+\tan^2(\frac{\theta_0}{2})\cos[2\varphi_0+
	\sigma(t)]}\right].
	\label{varphi1}
	\xeea
This notation allows one to directly read off the eigenvectors
$|n^{(1)};t\xkt$ and eigenvalues $E_{n^{(1)}}(t)$ of $H^{(1)}(t)$ by
replacing $n$ by $n^{(1)}$ and $(r,\theta,\varphi)$ by $(r^{(1)},
\theta^{(1)},\varphi^{(1)})$ in Eqs.~(\ref{eg-ve}) and (\ref{eg-va}).
Particularly interesting is the fact that $r^{(1)}(t)>0$, unless the
angular speed of the tip of the magnetic field vanishes. $\omega(t)
=0=r^{(1)}(t)$ define the times $t$ for which $H^{(1)}(t)=0$. These
``degenerate'' situations need further treatment. Here I suffice to
remind the reader that for time periods during which $H^{(1)}(t)=0$,
the adiabatic approximation is exact. 

Eqs.~(\ref{r1-De})-(\ref{varphi1}) together with (\ref{eg-ve}), (\ref{ui}),
and (\ref{al}) can be used to obtain $U^{(i+1)}(\tau)$ for all $i=0,1,
\cdots$. This is done by substituting $r^{(i)},~\theta^{(i)},~\varphi^{(i)}$
for $r,~\theta,~\varphi$ in Eqs.~(\ref{r1-De})-(\ref{varphi1}), (\ref{eg-ve}),
(\ref{w}),  (\ref{al}) and using (\ref{ui}). The result is:
	\xbea
	U^{(i+1)}(\tau)&=&W\left( \theta^{(i)}(\tau),\varphi^{(i)}(\tau)\right)
	\sum_n e^{i\alpha_{n^{(i)}}(\tau)}|n,(r_0,0,0)\xkt\xbr n,(r_0,0,0)|
	W\left( \theta^{(i)}(0),\varphi^{(i)}(0)\right)\;,\xnn\\
	&=&W\left( \theta^{(i)}(\tau),\varphi^{(i)}(\tau)\right)\:
	e^{i\frac{\alpha^{(i)}(\tau)}{\hbar}J^3}\:
	W\left( \theta^{(i)}(0),\varphi^{(i)}(0)\right)\;,
	\label{ui-1}
	\xeea
where $\alpha^{(i)}(t):=\delta^{(i)}(t)+\gamma^{(i)}(t)$ and
	\xbea
	\delta^{(i)}(t)&:=&-b\int_0^t r^{(i)}(t')\,dt'\,,~~~
	\gamma^{(i)}(t)\::=\:-\int_0^t [1-\cos\theta^{(i)}(t')]\,
	\dot\varphi^{(i)}(t')\,dt'\,,
	\label{al-de-ga-1}\\
	\alpha_{n^{(i)}}(t)&=&n\alpha^{(i)}(t)\,,~~~
	\delta_{n^{(i)}}(t)\:=\:n\delta^{(i)}(t)\,,~~~
	\gamma_{n^{(i)}}(t)\:=\:n\gamma^{(i)}(t)\,.
	\label{al-de-ga-n}
	\xeea

For $\varphi_0=0$ Eqs.~(\ref{r1-De})-(\ref{varphi1}) simplify considerably.
In this case one has:
	\xbe
	r^{(1)}(t)=\frac{\omega(t)}{b}\;,~~~
	\theta^{(1)}(t)=\cos^{-1}[-\sin\theta_0\cos\sigma(t)]\;,~~~
	\varphi^{(1)}(t)=-\tan^{-1}\left[ \frac{\tan\sigma(t)}{\cos\theta_0}
	\right]\;.
	\label{phi0=0}
	\xee
Substituting these relations in Eq.~(\ref{al-de-ga-1}) with $i=1$, 
and performing the necessary algebra, one finds:
	\xbea
	\delta^{(1)}(t)&=&-\int_0^t v(t')dt'=:-\ell(t)\,,~~~
	\gamma^{(1)}(t)\:=\:- \tan^{-1}\left[ \frac{X(t)+Y(t)}{1-X(t)Y(t)}\right]\;,
	\label{de-ga-2}\\
	X(t)&:=&-\frac{\cos\theta_0[\tan\sigma(t)-\tan\sigma(0)]
	 }{\cos^2\theta_0+\tan\sigma(t)\tan\sigma(0)}\;,~~~
	Y(t)\::=\:\tan\theta_0[\sin\sigma(t)-\sin\sigma(0)]\;,
	\label{xy}
	\xeea
where $\ell(t)$ is the length of the projection of the portion of  the 
curve $C$ which lies between  $C(0)$ and $C(t)$ onto the unit sphere
centered at $r=0$. The appearance of such a geometric quantity
in this calculation is quite remarkable.

Let us next apply the adiabatic product expansion to identify a
class of exactly solvable cases. This is done by enforcing
$H^{(N+1)}(\tau)=0$ or alternatively $r^{(N+1)}(\tau)=0$. The 
first nontrivial case is $N=1$, i.e., $r^{(2)}(\tau)=0$.  This
condition may be easily fulfilled by requiring $\dot\sigma=0$
which immediately leads to:
	\xbe
	r(t)=\frac{1}{b}\left[ \cos\theta\:\dot\varphi-
	\frac{  \frac{d}{dt}(\frac{\dot\theta}{\sin\theta\,\dot\varphi})}{
	1+(\frac{\dot\theta}{\sin\theta\,\dot\varphi})^2}\right]=
	\frac{\dot\varphi}{b}\left[\cos\theta-
	\frac{\frac{d}{d\varphi}(\frac{\theta'}{\sin\theta})}{
	1+(\frac{\theta'}{\sin\theta})^2}\right]\;,
	\label{r=}
	\xee
where $\theta'=d\theta/d\varphi$. Thus, for every $\theta=
\theta(\varphi)$ which renders the right hand side of 
(\ref{r=}) positive for all $t$, $|\psi(t)\xkt=U^{(0)}(t)U^{(1)}(t)
|\psi_0\xkt$ is the exact solution of the Schr\"odinger
equation (\ref{sch-eq}). Clearly, one can generate further
exactly solvable examples corresponding to $N=2,3,\cdots$,
in this manner.

The adiabatic product expansion suggested in this article may be
viewed as an iterative procedure to integrate the Schr\"odinger 
equation. It  involves only algebraic manipulations and integration
of  scalar functions. It also provides a generalization of the quantum 
adiabatic approximation which involves an integer $N$ as its order. 
The approximation yields the exact result if  $H^{(N+1)}(\tau)=0$. 
In the general case where $H^{(i)}(\tau)\neq 0$, for all $i$, one 
may improve the accuracy of the approximation indefinitely by
increasing its order.

As shown for the magnetic dipole system, one may be able to
obtain generic expressions for the terms in the adiabatic product
expansion. These can be used to generate exactly solvable
examples. In particular, if the dynamics of the system is given
by a dynamical group $G$, i.e., if the Hilbert space provides a
unitary irreducible representation $\rho$ of a Lie group $G$ 
and $H(t)=\rho[h(t)]^\dagger H_d(t)\rho[h(t)]$, with $h(t)\in G$ 
and $H_d(t)$ diagonal, belongs to the representation of the 
Lie algebra of $G$ and $U(t)=\rho[g(t)]$ for some $g(t)\in G$ 
\cite{p7}, then one might try to employ a similar approach as the
one used in the analysis of the dipole system ($G=SU(2)$).  In
particular this approach may be readily applied for the generalized
harmonic oscillator  system where  $G=SU(1,1)$, \cite{g-h-o}.
An interesting direction for further study of the adiabatic product
expansion is to seek its meaning and implications in the path 
integral approach to quantum mechanics and ultimately quantum 
field theory.

\section*{Acknowledgements}
I would like to thank Bahman Darian and Ertu$\check{\rm g}$rul
Demircan for reading the first draft of this article. I would also like
to acknowledge the financial support of the Killam Foundation of
Canada. 

\end{document}